# Multimedia Description Framework (MDF) for Content Description of Audio/Video Documents

Michael J. Hu          Ye Jian
School of EEE, Nanyang Technological University
Singapore 639798

**[Abstract]**      MPEG is undertaking a new initiative to standardize content description of audio and video data/documents. When it is finalized in 2001, MPEG-7 is expected to provide standardized description schemes for concise and unambiguous content description of data/documents of complex media types. Meanwhile, other meta-data or description schemes, such as Dublin Core, XML, etc., are becoming popular in different application domains. In this paper, we propose the Multimedia Description Framework (MDF), which is designated to accommodate multiple description (meta-data) schemes, both MPEG-7 and non-MPEG-7, into integrated architecture. We will use examples to show how MDF description makes use of combined strength of different description schemes to enhance its expression power and flexibility.  We conclude the paper with discussion of using MDF description of a movie video to search/retrieve required scene clips from the movie, on the MDF prototype system we have implemented.

## 1.      INTRODUCTION

There has been growing interest worldwide in providing content description of varied formats in different application domains. It is suggested that concise and proficient description of a large collection of text documents, music records, news/documentaries, movie videos,  image data banks, etc. could help users in fast search and retrieval of data/documents of numerous media types in digital libraries, multimedia-on-demand (MoD) on the Internet, multimedia production, tele-education, etc.. With such background, Motion Picture Expert Group (MPEG) has recently initiated a new standardization project, called as **MPEG-7**.

MPEG-7 is formally named as "**Multimedia Content Description Interface**". It will extend the limited capabilities of current solutions in identifying multimedia content, by including more data types and specifying a standard set of description schemes that can be used to describe various types of multimedia information. MPEG-7 will also standardize ways to define and extend different description schemes and their relationships. Such description will then be used to support fast and efficient search of audio/video data/documents such as pictures, graphics, audio, speech, and video [5][6][7].

Taking this broad range of requirements into account, MPEG-7 will not define a monolithic system for content description, but a set of methods and tools for different levels of multimedia description [5][6]. In particular, it will standardize following elements, which are applicable for concise and proficient content description of different types of multimedia data in a wide range of applications:

1) **A set of description schemes (DSs) and descriptors (Ds)** for content description of audio records, speech clips, movies, video documentaries, etc.;

2) **Description Definition Language (DDL)** that provides standardized grammar and syntax for defining and denoting DSs and Ds, so that concise and unambiguous content description can be parsed by machines, and shared by different systems for interoperability;

3) **A scheme for coding the description**.

Meanwhile, other description (or meta-data) schemes have been emerging. Some of them have been enthusiastically accepted worldwide in diversified application domains. For instance, the Dublin Core meta-data scheme is widely used for simple description, such as author names, date of publication, etc., of libraries' collection. Others, such as XML [1] and multimedia extension of Dublin Core [3], have been under development as meta-data or description schemes for applications involving web pages and Internet resources, and complex media data (e.g., image/audio) respectively.

We project that different meta-data or description schemes, MPEG-7 or non-MPEG-7, will co-exist for a long time, as each of these schemes may have its own strength of describing a particular type of data/document in a specific domain. We further suggest that blending different description (meta-data) schemes into integrated description framework/architecture should result in more powerful and flexible description. Warwick Framework [31] and Resource Description Framework (RDF) [30] are two recent examples of such framework/architecture.

In this paper, we propose the Multimedia Description Framework (MDF), which is designated to accommodate different multimedia description schemes, both MPEG-7 and non-MPEG-7, in powerful and flexible description of multimedia data. We will use examples to demonstrate benefit and flexibility of using multiple description schemes and their combined expressing power in MDF description. Our discussion in the paper is organized as follows. First, in Section 2, we present MDF, its syntax and basic structure; we also discuss advantages and benefits of using such framework. Thereafter, we discuss in Section 3 our implementation of a Java-based MDF prototype system. We are currently running and testing the implemented system with data provided by MPEG-7 [33]. A comprehensive example of MDF description of a movie video, "Ultra-steel suspends the longest bridge in the world", is given in Section 4 and Appendix II. Finally, our further research direction will be pointed out before we reach our conclusion in Section 5.

## 2. MULTIMEDIA DESCRIPTION FRAMEWORK (MDF)

Multimedia Description Framework (MDF) can be formally denoted as a three-tuple,

$$(E, P, VP)$$

in which **E** stands for the entity (e.g., an image, a video document, etc.) that is to be described, **P** represents a set of properties used to describe the entity, and **VP** is the viewpoint in which such description is made.

At a specific viewpoint, an entity can be described by a set of well-defined properties specifying particular aspects, characteristics, attributes of the entity, as well as its relationship to other entities. For instance, Figure 1 shows a simple example that property (type) of the entity (a document, pointed by its URI: http://155.69.66.134:8000/Nhkvideo.mpg) is multimedia. Other properties of the entity, such as its creator of the document, its creation date, title, summary, etc., can also be represented in similar format.

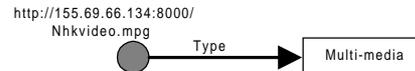

**Figure 1    Entity and properties of the entity**

However, at different viewpoints, different sets of properties may be needed to describe the same entity. For example, an entity may look different at different granularity levels, and therefore, should be described by different properties. Figure 2 illustrates that, when the same entity is described at different viewpoints (i.e. Viewpoint 1, Viewpoint 2, …), different sets of property should be applicable, describing the entity at various details.

We will show in the rest of this paper how Multimedia Description Framework (MDF) is designated to accommodate multiple and different viewpoints, or description scheme, in comprehensive MDF description. We will also use examples to demonstrate how this effectively enhances the expressive power and flexibility of such description.

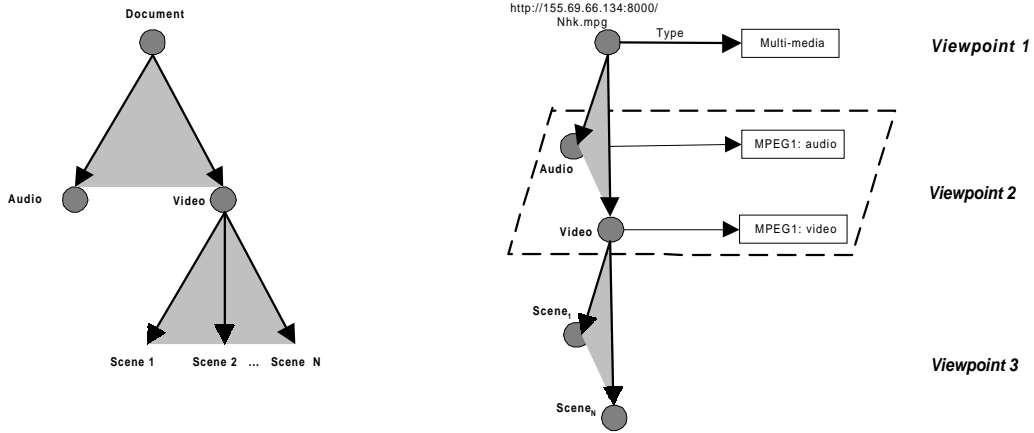

**(a) Hierarchy structure of a video document**

**(b) Describing the video document at different granularity/viewpoint**

**Figure 2     MDF with different viewpoints**

## 2.1     MDF STRUCTURE

The fundamental structure of MDF description includes following three elements:

1) **MDF syntax declaration**:  Syntax and grammar used in the description.

2) **MDF semantics declaration**:  Viewpoint(s) used in the description. The notation used for viewpoint names is

   **VP:VIEWPOINT_NAME**, e.g., *VP:DOC*

3) **MDF description**: the description and the entity it describes.

A brief summary of MDF structure is given in Table 1, whereas a simple example of MDF description is given on the next page. In the example, MD$^2$L is used as MDF syntax . The viewpoint used in the description is *VP:DOC*, its schema is defined in *http://155.69.66.134:8000/ DOC*. Thereafter, an MDF description is given, describing properties (e.g., title, subject, etc.) of the a movie video, http://155.69.66.134:8000/doc /Nhkvideo.mpg: the title of the document is "`Ultra-steel suspends the longest bridge in the world`", etc.

| | | |
|---|---|---|
| ***MDF*** | *:: =* | ***['<MDF:MDF' S 'SYNTAX' EQ '"'URI-reference '"'S '>']*** <br> ***VpDecl description\* ['</MDF:MDF]***  **(2.1)** |
| ***vpDecl*** | *:: =* | ***'<?MDF VP:' vpName 'href=' S '"' URI-reference '"' S '?>'*** <br> **(2.2)** |
| ***description*** | *:: =* | ***'<MDF:Description aboutAttr? '>' propertyElement\**** <br> ***'</MDF:Description>'***  **(2.3)** |
| ***aboutAttr*** | *:: =* | ***'About = ' ' URI-reference ' " '***  **(2.4)** |
| ***propertyElement*** | *:: =* | ***'<' propertyName '>' propertyValue '</' propertyName '>'*** <br> **(2.5)** |
| ***vpName*** | *:: =* | ***'VP:' Qname***  **(2.6)** |
| ***propertyName*** | *:: =* | ***Qname***  **(2.7)** |
| ***propertyValue*** | *:: =* | ***description \| string***  **(2.8)** |

**Table 1     Formal notation of MDF description**

```
<MDF:MDF  SYNTAX= "http://155.69.66.134:8000/MDL/ " >
<?MDF  VP:DOC  href="http://155.69.66.134:8000/DOC" ?>
        <MDF:Description  About= " http://155.69.66.134:8000/doc/Nhkvideo.mpg">
                <DOC:Title> Ultra-steel suspends the longest bridge in the world</DOC:Title>
                <DOC:Subject>  Ultra-steel, bridge construction </DOC:Subject>
                                ...
        </MDF:Description>
</MDF:MDF>
```

## 2.2    MDF SYNTAX:  MD$^2$L

Unless it is explicitly declared, the default syntax of MDF is the Multimedia Description Definition Language (MD$^2$L) developed by our team in the School of Electrical & Electronic Engineering in Nanyang Technological University [19].  MD$^2$L was designated to meet requirements of MPEG-7 DDL [5][6].

MD$^2$L provides a formal notation for presenting description schemes and descriptors designated for content description of various types of media data/document such as audio records, video clips, etc.  Key elements involved in MD$^2$L are:

1)  **MD$^2$L Declaration**        It gives general information about MD$^2$L syntax options and grammar,  and also specifies what a description scheme should include. An MD$^2$L parser may allow the declaration to be omitted.  In those circumstances,  standard default settings (called the **Reference MD$^2$L Syntax** [19]) are used.

2)  **Description Scheme Declaration (DSD)**  It defines what descriptors, or other description

schemes, are admitted in the description set, how different DSs and their respective descriptors are related to one another, etc. Each MPEG-7 DS represents an independent viewpoint in MDF.

3)  **A Description Instance**   It illustrates how a set of descriptors can collectively formulate a set of valid description, by giving a concrete example of using declared DSD.

With MD$^2$L, different description schemes could be organized into **DS Hierarchy**. For instance, Figure 3 shows an example DS hierarchy that we are currently using in our prototype MDF system.  In such hierarchy, **DOC** specifies the description scheme that is universally applicable to all types of media data/documents including text documents, image files, and audio/video streams. ***TEXT***, ***IMAGE***, ***AUDIO*** and ***VIDEO*** are the **first-generation** DSs, and therefore, children of *DOC*.   Each of them specifies descriptors that are only applicable to the specific type of document/data. Meanwhile, they inherit specification of descriptors from its higher-layer DS, i.e., *DOC*, and so on.

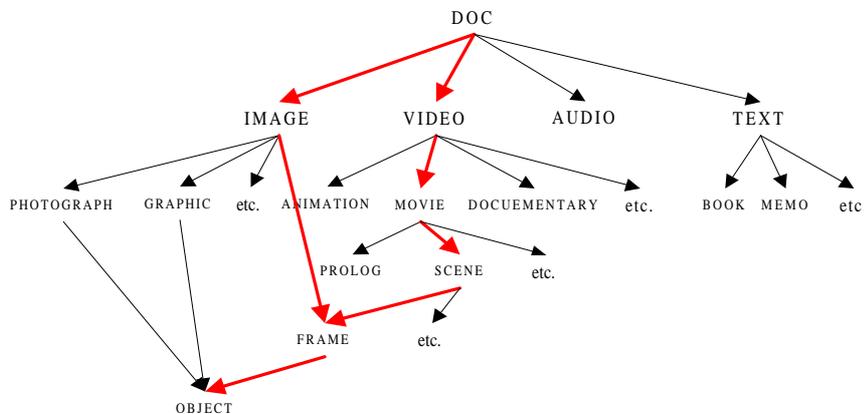

**Figure 3      DS hierarchy for data/documents of different media types**

## 2.3    EXAMPLES OF USING MDF

With MD$^2$L as its syntax, MDF description of all sorts of media data/document can be worked out. Following examples are MDF descriptions of the same movie video at different viewpoints, including *VP:DOC, VP:VIDEO*, *VP:MOVIE*, and *VP:SCENE*. Viewpoints (description schemes) used are explicitly specified at the beginning of each MDF description. Definition of respective DSs (*DOC, VIDEO, MOVIE, SCENE, FRAME, OBJECT*) of Figure 3 is given in the Appendix I.

### EXAMPLE 1:  A DOCUMENT — *VP:DOC*

```
<MDF:MDF SYNTAX= "http://155.69.66.134:8000/MDL/ " >
<?MDF VP:DOC href="http://155.69.66.134:8000/DOC" ?>

        <MDF:Description About= " http://155.69.66.134:8000/doc/Nhkvideo.mpg">
                <DOC:Title> Ultra-steel suspends the longest bridge in the world </DOC:Title>
                <DOC:Subject> Ultra-steel, bridge construction </DOC:Subject>
                <DOC:Publisher>NHK </DOC:Publisher>
                <DOC:Type> Doc</DOC:Type>
                <DOC:Description> This movie is a scientific film about how Ultra-steel invented and what
                characteristics it has to support the construction of the longest bridge in the world
                </DOC:Description>
                <DOC:Format> mpeg-1</DOC:Format>
        </MDF:Description>
</MDF:MDF>
```

### EXAMPLE 2:  A VIDEO — *VP:VIDEO*

```
<MDF:MDF SYNTAX= "http://155.69.66.134:8000/MDL/ " >
<?MDF VP:VIDEO  href="http://155.69.66.134:8000/VIDEO" ?>

        <MDF:Description  About= " http://155.69.66.134:8000/doc/Nhkvideo.mpg">
                <!---Descriptors inherit from DOC DS--!>
                <VIDEO:Title> Ultra-steel suspends the longest bridge in the world </VIDEO:Title>
                <VIDEO:Subject> Ultra-steel, bridge construction </VIDEO:Subject>
                <VIDEO:Publisher> NHK</VIDEO:Publisher>
                <VIDEO:Type> Video </VIDEO:Type>
                <VIDEO:Description> This movie is a scientific film about how Ultra-steel invented and what
                characteristics it has to support the construction of the longest bridge in the world.
                </VIDEO:Description>
                <VIDEO:Format> mpeg-1 </VIDEO:Format>
                <!—additional descriptors from VIDEO DS--!>
                <VIDEO:Duration> 00:09:04:15</VIDEO:Duration>
                <VIDEO:Rate> 30</VIDEO:Rate>
        </MDF:Description>
</MDF:MDF>
```

### EXAMPLE 3:  A MOVIE — *VP:MOVIE*

```
<MDF:MDF SYNTAX= "http://155.69.66.134:8000/MDL/ " >
<?MDF VP:MOVIE href="http://155.69.66.134:8000/MOVIE" ?>

        <MDF:Description  About= " http://155.69.66.134:8000/doc/Nhkvideo.mpg">
```

```
            <!---Descriptors inherit from VIDEO DS--!>
            <MOVIE:Title> Ultra-steel suspends the longest bridge in the world </MOVIE:Title>
            <MOVIE:Subject> Ultra-steel, bridge construction </MOVIE:Subject>
            <MOVIE:Publisher> NHK</MOVIE:Publisher>
            <MOVIE:Type> Movie </MOVIE:Type>
            <MOVIE:Description> This movie is a scientific film about how Ultra-steel invented and what
            characteristics it has to support the construction of the longest bridge in the world.
            </MOVIE:Description>
            <MOVIE:Format> mpeg-1 </MOVIE:Format>
            <MOVIE:Duration>00:09:04:15</MOVIE:Duration>
            <MOVIE:Rate> 30 </MOVIE:Rate>
            <!—additional descriptors MOVIE DS--!>
            <MOVIE:Cast>Reporter, Tina Kawamura</MOVIE:Cast>
            <MOVIE:Cast>Research Engineer, Dr. Toshihiko Takahashi</MOVIE:Cast>
            <MOVIE:Movietype> scientific</MOVIE:Movietype>
    </MDF:Description>
</MDF:MDF>
```

**EXAMPLE 4:   A SCENE --*VP:SCENE***

```
<MDF:MDF SYNTAX= "http://155.69.66.134:8000/MDL/ " >
<?MDF VP:SCENE href="http://155.69.66.134:8000/SCENE" ?>

        <MDF:Description About= " http://155.69.66.134:8000/doc/Nhkvideo.mpg#scene1">
            <!---Descriptors inherit from MOVIE DS--!>
            <SCENE:Title> Ultra-steel suspends the longest bridge in the world </SCENE:Title>
            <SCENE:Subject> Ultra-steel, bridge construction</SCENE:Subject>
            <SCENE:Publisher> NHK  </SCENE:Publisher>
            <SCENE:Type> Scene </SCENE:Type>
            <SCENE:Description> Air-shot of the Akashi-channel bridge </SCENE:Description>
            <SCENE:Format> mpeg-1 </SCENE:Format>
            <SCENE:Duration>  00:00:08:13 </SCENE:Duration>
            <SCENE:Rate> 30</SCENE:Rate>
            <SCENE:Cast> Reporter, Tina Kawamura </SCENE:Cast>
            <SCENE:Cast>Research Engineer, Dr. Toshihiko Takahashi</SCENE:Cast>
            <SCENE:Movietype> scientific</SCENE:Movietype>
            <!--additional descriptors from SCENE DS--!>
            <SCENE:Keyframe> http://155.69.66.134:8000/Nhkvideo.mpg#frame1</SCENE:Keyframe>
            <SCENE:StartTime>00:00:00:00</SCENE:StartTime>
            <SCENE:EndTime>00:00:08:13</SCENE:EndTime>
            <SCENE:Camera_Distance>AIR-SHOT</SCENE:Camera_Distance>
            <SCENE:Object> http://155.69.66.134:8000/movie/bridge.mdf </SCENE:Object>
    </MDF:Description>
</MDF:MDF>
```

## 2.4    ADVANTAGE OF USING MDF

Examples given above provide basic description of entities (e.g., a movie, a scene, etc.) at individual viewpoints. Such description is static, and it may not always provide the flexibility one needs. The proposed solution is to use multiple description schemes in MDF description. For instance, if we may want to describe the movie video at different levels of detail (Example 5), we could simply import several description schemes (e.g., *VP:MOVIE*, *VP:SCENE*, etc.) into an MDF description. As the result, descriptors from all of these schemes can be used. This certainly increases the expressing power and flexibility of the MDF description.

**Example 5        MULTIPLE VIEWPOINTS -- *VP:MOVIE, VP:SCENE, & VP:OBJECT***

```
<MDF:MDF SYNTAX= "http://155.69.66.134:8000/MDL/ " >
<?MDF VP:MOVIE  href="http://155.69.66.134:8000/VIDEO" ?>
<?MDF VP:SCENE  href="http://155.69.66.134:8000/SCENE" ?>
<?MDF VP:OBJECT  href =" http://155.69.66.134:8000/OBJECT" ?>

        <MDF:Description  About= " http://155.69.66.134:8000/doc/Nhkvideo.mpg"
                <!---Descriptors inherit from VIDEO DS--!>
                <MOVIE:Title> Ultra-steel suspends the longest bridge in the world </MOVIE:Title>
                <MOVIE:Movietype> scientific</MOVIE:Movietype>
                <MOVIE:Scene>
                        <SCENE:ID>SCENE1</SCENE:ID>
                        <SCENE:Keyframe>http://155.69.66.134:8000/Nhkvideo.mpg#frame1</SCENE:Keyfra
                        me>
                        <SCENE:Object>
                                <OBJECT:ID>...</OBJECT:ID>
                        </SCENE:Object>
                </MOVIE:Scene>
                <MOVIE:Scene>
                        <SCENE:ID>SCENE2</SCENE:ID>
                </MOVIE:Scene>
        </MDF:Description>
</MDF:MDF>
```

Another advantage of introducing MDF is that we project there will always be multiple viewpoints, or description schemes, available to describe varied types of data/documents. Some of them will be accepted as standard MPEG-7 description schemes (DSs) when MPEG-7 is finalized in 2001. Others may be preferred in different application domains such as digital libraries and applications involving WWW pages/Internet resource, or proved to be more suitable for specific data types such as text or music records.  More are yet to emerge, as technology becomes more sophisticated and developed (Table 2).

| Standard MPEG7 DS For audio/video documents | | ▪ This is still under proposal,  and will be finalized in 2001. ▪ The hierarchical DS structure of Figure 4 is an example of MPEG7 description scheme, which we are proposing to the MPEG7 committee. |
|---|---|---|
| Non-MPEG7 Description scheme | Dublin Core (DC) for text | ▪ 15 elements that are widely used in libraries for bibliographical information and description; |
| | Dublin Core Extension for Multimedia data | ▪ Under development [2][32] |
| | XML/RDF Description scheme | ▪ Meta-data, or description scheme, for web-based information and Internet resources |
| | Other new description scheme | ▪ To be emerging in the future. |

**Table 2  Different viewpoints and description schemes**

With the proposed framework, different description schemes (MPEG-7 or non-MPEG-7) can be combined into integrated MDF description. This enhances flexibility and expressing power of MDF description. We use Examples 6 and 7 to show how different description schemes can be declared and applied concurrently. In Example 6, two viewpoints, **VP:VIDEO** and **VP:OCLC**, are used in content description of a movie video. The former provides MPEG-7 description scheme for the video, whereas the latter is used to denote descriptors (e.g. ISBN number) popularly used in the libraries, or video-rent/book stores. In Example 7, two viewpoints (**VP:VIDEO** and **VP:AUDIO**) are used to describe the same movie video that is composed of both video sequence and audio/music track. Description of both video sequence and audio track would help users to search/retrieve the item.

**EXAMPLE 6:  MULTIPLE VIEWPOINTS -- *VP:VIDEO* & *VP:OCLC***

*<MDF:MDF  href =" http://155.69.66.134:8000/MDL/" >*
*<?MDF VP:VIDEO  href =" http://155.69.66.134:8000/VIDEO/"  ?>*
*<?MDF VP:OCLC    href =" http://naa.gov.au/OCLC#"  ?>*

        *<MDF:Description  About= " http://155.69.66.134:8000/doc/Nhkvideo.mpg">*
           ***<!---Descriptors inherit from DOC DS--!>***
           *<VIDEO:Title> Ultra-steel suspends the longest bridge in the world </VIDEO:Title>*
           *<VIDEO:Subject> Ultra-steel, bridge construction </VIDEO:Subject>*
           *<VIDEO:Rate> 30</VIDEO:Rate>*
           *...*
           *<OCLC:ISBN>  981-3076-75-99 </OCLC:ISBN>*
        ***</MDF:Description>***
*</MDF:MDF>*

**EXAMPLE 7: MULTIPLE VIEWPOINTS -- *VP:VIDEO* & *VP:AUDIO***

*<MDF:MDF  href =" http://155.69.66.134:8000/MDL/">*
*<?MDF VP:VIDEO  href =" http://155.69.66.134:8000/VIDEO/"  ?>*
*<?MDF VP:AUDIO  href =" http://155.69.66.134:8000/AUDIO/"  ?>*

        *<MDF:Description  About= " http://155.69.66.134:8000/doc/Nhkvideo.mpg">*
           ***<!---Descriptors inherit from DOC DS--!>***
           *<VIDEO:Title> Ultra-steel suspends the longest bridge in the world </VIDEO:Title>*
           *<VIDEO:Subject> Ultra-steel, bridge construction </VIDEO:Subject>*
           *<VIDEO:Rate> 30 </VIDEO:Rate>*
           *...*
           *<AUDIO:Rate>... </AUDIO:Rate>*
           *<AUDIO:Mode> Stereo </AUDIO:Mode>*
        ***</MDF:Description>***
*</MDF:MDF>*

## 2.5    OTHER RELATED DESCRIPTION FRAMEWORKS

As a description framework,   MDF extends $MD^2L$ in the same way as the Resource Description Framework (RDF) does to XML, and Warwick Framework to Dublin Core.

**RESOURCE DESCRIPTION FRAMEWORK**

The Resource Description Framework (RDF) – developed by the World Wide Web Consortium (W3C) – provides the foundation for meta-data interoperability across different resource description communities. As it was pointed out in [30], "One of the major obstacles facing the resource description community is the multiplicity of incompatible standards for meta-data syntax and schema definition languages. This has lead to the lack of, and low deployment of, cross-discipline applications and services for the resource description communities." RDF provides a solution to these problems by introducing different schemes, or name spaces, via syntax and schema specifications.

The objective of RDF is to support the interoperability of meta-data. It allows descriptions of Web resources – any objects with a Uniform Resource Identifier (URI) as its address – to be made available in machine understandable form, enabling semantic of objects to be expressible and exploitable and enhancing applications of such resources.

As MDF and RDF are focused on different application domains, their syntax and schema specifications then are quite different. 1) RDF syntax is based on XML, and its schema specification involves mainly various meta-data schemes (e.g., Dublin Core, AGRL, etc.). The syntax specification of MDF is the Multimedia Description Definition Language (MD$^2$L), which is designated for complex media data/document such as image, audio, and video. MDF semantic scheme is also much more complicated than that of RDF, and in fact, it constitutes the superset of the latter. This is because description schemes used for complex media data/document is much more complicated as it is shown in Sections 2 and 3 of this paper. 2) RDF uses RDF schema to define a collection of namespaces, and due to the interoperability of these schema, it is able to support reusability of metadata. Similarly, MDF imports DSDs as its schema and makes use of other existing meta-data schema as viewpoints. The difference between MD$^2$L's DSD and XML's DTD is that DTD only gives specific constraints on structure of a document, while DSD offers function of making use of existed DSD, inheritance and partially inheritance of descriptors from parents. Moreover, MDF and DSD offer an unique identifier to identify every descriptor in order to void re-definition of existed descriptors. This makes defining DS more flexible.

## WARWICK FRAMEWORK

Another important architecture is the Warwick Framework, which was proposed and developed as "a container architecture" for Dublin Core and other existing, more specialized descriptive systems such as library cataloging (AACR 2/MARC) and the FGDC Meta-data scheme [31]. Similarly to RDF, Warwick Framework is to provide a more concrete and operationally useable formulation of the Dublin Core, in order to promote greater interoperability among content providers, content catalogers and indexers, and automated resource discovery/ description systems.

## 3 USING MDF IN MULTIMEDIA APPLICATION

Figure 4 shows the workflow, in which an inputted MDF description is checked by an MDF parser, and the confirmed MDF description is subsequently processed and indexed before it can be searched, accessed and retrieved. We have designed and implemented three key modules, including parsing, indexing, and searching modules, in a prototype system (Figure 5). These modules would be elaborated in this section.

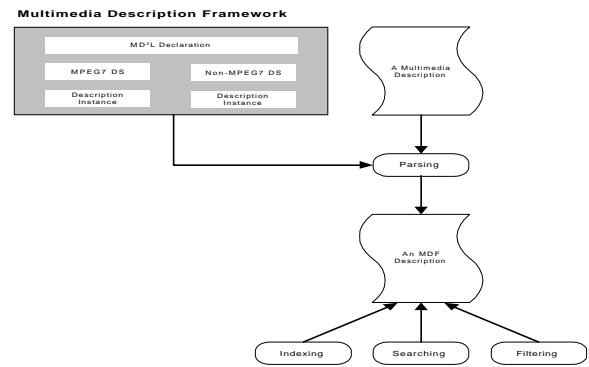

**Figure 4    Parsing and applying MDF description**

## 3.1    SYSTEM ARCHITECTURE

As it is shown in Figure 5, our system includes following key elements: a multimedia database system with multimedia data/document collections and its content description, an auxiliary index database (index of descriptors), a parsing module, an indexing module, and a searching/retrieving module. Our system is implemented with Java and Java Servlet. Whilst most of the system modules run on the server, the index and searching/retrieving modules can be accessed from client side (Section 3.5).

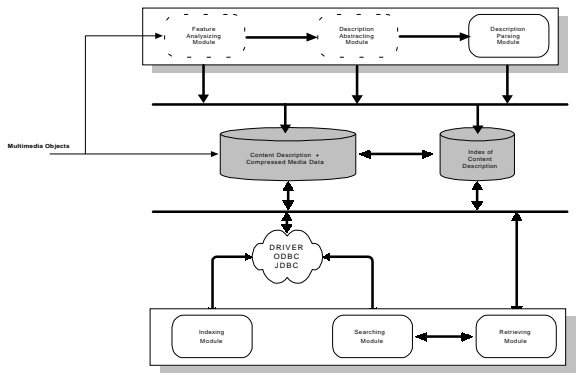

**Figure 5    System architecture**

It should be noted that automatic feature analysis and description abstraction for complex media data/ documents such as audio and video are challenging issues that under investigation by many researchers worldwide. Manual operation is currently involved in formulating the content description in our prototype system (in dashed boxes in Figure 5).

## 3.2    MULTIMEDIA DATABASE

In our system, an MDF description is normally stored along with the data/document it describes. If they are separated, the document URL should be explicitly specified in MDF Description.

Our system works as follows: 1) When multimedia data/objects are loaded into the system, the compressed media objects are inputted directly in the database; 2) Meanwhile, these objects are processed for feature analysis and description abstraction;  3) The abstracted description is further checked in MDF parsing, and the confirmed MDF description will be stored along with the multimedia data/objects for further processing and search/retrieval.

## 3.3    PARSING MODULE

The MDF parser we have implemented checks 1) grammar and syntax of MDF description, i.e., $MD^2L$; and 2) semantics (DSDs) used in MDF description. Main functionality of the MDF parser includes:

- Parsing MDF description, and reporting any errors found in the description;
- Generating a hierarchy tree for the description;
- Extracting key descriptors from MDF description for indexing.

Parsing the MDF description involves checking and confirming if the syntax used in MDF description conforms to $MD^2L$, or other alternatives declared in MDF syntax statement. The DSD hierarchy has to be formulated, in order to check and confirm that all element types used in the description are acceptable and valid (Figure 6). If no syntax and grammar errors reported by MDF parser, it generates DOM tree that can be further processed.

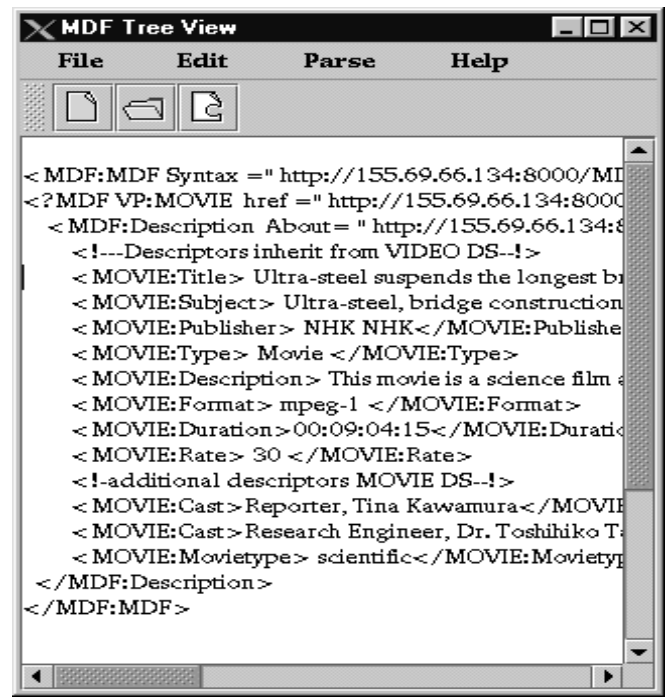

**Figure 6   Interface of MDF parser**

### 3.4    INDEXING KEY DESCRIPTORS

Key descriptors of an MDF description are abstracted, and subsequently, indices of these descriptors are formulated in our prototype system. For instance, following descriptors from *DOC* are indexed in our implemented prototype system: subject and keywords, title, description, author, format, etc.

The target of indexing module is to automatically formulate indices of key descriptors, which can be used to support efficient search and retrieval of requested multimedia data and documents. The conventional indexing algorithm is used in our current implementation, resulting in a number of inverted files as indices of key descriptors, which are stored in auxiliary index database.

### 3.5    USING CONTENT DESCRIPTION FOR QUICK SEARCH/RETRIEVAL

The search/retrieval module locates items whose description marches users' requests, by searching individual indice(s) specified by users. It returns the IDs/locations of related multimedia document(s) (which contains the requested content). These IDs/locations can then be followed to retrieve the respective multimedia data or documents. It's the ultimate goal of MPEG-7 standardization, that fast and real-time search of multimedia content description would lead to more efficient retrieval of respective multimedia documents [5].

Figures 7 show a typical application scenario, in which a user wants to search for any *video* documents that "*Ultra steel*" is included in its *keyword* description. The user submits his requests for "video", with subject/keyword "Ultra steel" to our search engine (Figure 7), and a list of multimedia documents in which these two terms are included in the content description (i.e., Subject/keyword) is found and displayed on the Interface. Retrieving one of these documents (by clicking one of returned documents, e.g., `155.69.66.134:8000/doc /Nhkvideo.mpg`) is illustrated in  Figure 7.

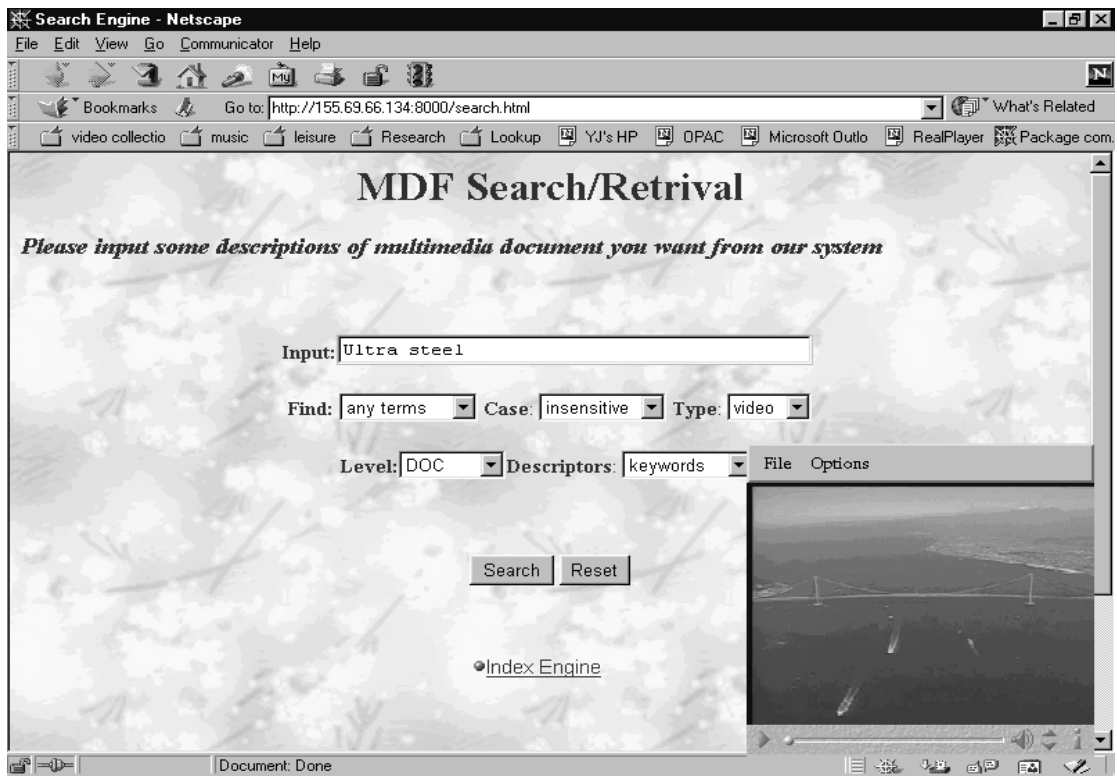

**Figure 7        Search/retrieval results of video with "Ultra steel"**

## 4 MDF DESCRIPTION: MOVIE VIDEOS

To illustrate its flexibility and effectiveness, we have been working on detailed MDF description of a large amount of multimedia data (MPEG-7 Content Set, in 33 CD-ROMs [33]). An example video clip from the set, " Ultra steel ", is used as the example in this paper, and detailed information about this video clip is given in APPENDIX II. We have also been using our implemented prototype system (Section 3) to parse and test the formulated MDF description. The confirmed MDF description will be applicable for fast content-based search/retrieval of required document(s) in our prototype system.

## 5. CONCLUSION

A new endeavor undertaken by the Motion Picture Expert Group (MPEG) is to standardize content description of different types of media data, especially those of complex types such as audio and video. Meanwhile, we note that several other meta-data or description schemes, such as Dublin Core, XML/RDF, etc., have been developed for different types of data and application. In this paper, we propose Multimedia Description Framework (MDF), which embraces multiple description (meta-data) schemes, both MPEG-7 and non-MPEG-7 schemes, into comprehensive description architecture. We have shown in this paper how MDF description can make use of combined strength of different description schemes, and therefore, enhances its expression power and flexibility. Some discussion of our implemented prototype system, and our ongoing test of formulating content description of a large amount of multimedia data in our prototype system, is also given in the paper.

# APPENDIX I       LIST OF DSD

*DOC DSD*

```
<? MDL  version = "1.0" ?>
<!DSD DOC [
<!DS DOC (Title , Subj , Author, Pub , Agent , Date, Res , Format , Identifier, Rel , Source, Language , Cov ,
Content, Rm) >
<!ATTLIST DOC  " id    ID      #REQUIRED">
<!D Title                       (#PCDATA)>
<!D Subj                        (#PCDATA)>
<!D Author                      (#PCDATA)>
<!D Pub                         (#PCDATA)>
<!D Agent                       (#PCDATA)>
<!D Date                        (#DATE)>
<!D Res                         (#PCDATA)>
<!D Format                      (#PCDATA)>
<!D Identifier                  (#URI)>
<!D Rel                         (#PCDATA)>
<!D Source                      (#PCDATA)>
<!D Language                    (#PCDATA)>
<!D Cov                         (#PCDATA)>
<!D Content                     (#PCDATA)>
<!D Rm                          (#PCDATA)>
] >
```

*IMAGE DSD*

```
<?MDL version = "1.0" ?>
<!DSD IMAGE [
<!DS  IMAGE (Shape, Sketch , Color, Visual, Texture, Motion, Effect, Spatial, Histogram) >
<!ATTLIST "id   ID    #REQUIRED" >
<!D Shape                       (#PCDATA)>  <!-- Under consideration--!>
<!D Sketch                      (#PCDATA) >  <!-- Under consideration--!>
<!D Color                       (#COMPOSITE)> <!-- Under consideration--!>
<!D Visual                      (#PCDATA)> <!-- Under consideration--!>
<!D Texture                     (#PCDATA) > <!-- Under consideration--!>
<!D Motion                      (#COMPOSITE)> <!-- Under consideration--!>
<!D Effect                      (#PCDATA)> <!-- Under consideration--!>
<!D Spatial                     (#PCDATA) >  <!-- Under consideration--!>
<!D Histogram                   (#PCDATA)> <!-- Under consideration--!>
 ] >
```

*VIDEO DSD*

```
<?MDL version = "1.0" ?>
<!DSD VIDEO  [
<!DS VIDEO ( Duration , Rate ) >
<!ATTLIST  VIDEO "id   ID   #REQUIRED">       <!--ID attribute for Video element -->
<!D Duration     (#TIME)>                     <!-- duration time of video -->
<!D Rate         (#FLOAT)>                      <!-- how many frames a second -->
] >
```

***MOVIE DSD***

```
<?MDL version = "1.0" ?>
<!DSD MOVIE   Parent = "VIDEO; http://155.69.66.134:8000/DSD/video.dsd"
                  Children = "SCENE" [
<!DS MOVIE ( Cast *, Director, Movietype, Scene* ) >
<!ATTLIST MOVIE "id    ID       #REQUIRED" >
<!D Cast                       (#PCDATA)>
<!D Director                   (#PCDATA)>
<!D Movietype                  (#PCDATA)>
<!DS Scene        "http://155.69.66.134:8000/DSD/scene.dsd">
]>
```

***SCENE DSD***

```
<?MDL version = "1.0" ?>
<!DSD SCENE  Parent  = "MOVIE; http://155.69.66.134:8000/DSD/movie.dsd" [
<!DS  SCENE (Keyframe *, Camera, StartTime, EndTime, Object*)>
<!ATTLIST  SCENE "id       ID    #REQUIRED">
<!DS Keyframe   "http://155.69.66.134:8000/DSD/frame.dsd">
<!D Camera_Distance           (#FLOAT)>
<!D Camera_Angle              (#FLOAT)>
<!D Camera_Motion             (#PCDATA)>
<!D StartTime           (#TIME)>      <!—seconds , frames no -- !>
<!D EndTime             (#TIME)>      <!—seconds , frames no -- !>
<!DS Object                "http://155.69.66.134:8000/DSD/object.dsd">
]>
```

***FRAME DSD***

```
<?MDL version = "1.0" ?>
<!DSD FRAME  Parent  = "SCENE; http://155.69.66.134:8000/DSD/scene.dsd"
                  Parent = "IMAGE; http://155.69.66.134:8000/DSD/image.dsd" [
<!DS  FRAME (Timestamp, Object*, Background, Color)>
<!ATTLIST  FRAME "id      ID      #REQUIRED">
<!D Timestamp           (#TIME) >     <!—frame no --!>
<!D Background          (#PCDATA)>          <!-- Under consideration--!>
<!DS Object              "http://155.69.66.134:8000/DSD/object.dsd">
<!D Color               (#COMPOSITE)>        <!-- Under consideration--!>
]>
```

***OBJECT DSD***

```
<?MDL version = "1.0" ?>
<!DSD OBJECT  Parent = "FRAME; http://155.69.66.134:8000/DSD/frame.dsd" [
<!DS  Object (Position, Shape, Motion, Spatial, temporal, Entity, Color)>
<!ATTLIST  Object "id    ID     #REQUIRED">
<!D Position      (#ARRAY)>
<!D Shape         (#PCDATA)>     <!-- Under consideration--!>
<!D Motion        (#COMPOSITE)>        <!-- Under consideration--!>
<!D Spatial       (#PCDATA)>     <!-- Under consideration--!>
<!D Entity        (#BLOB)>
<!D Color         (#PCDATA)>     <!-- Under consideration--!>
]>
```

[Note:   Binary Large Objects, BLOBs, are streams of bytes of arbitrary value and length.]

**APPENDIX II    DESCRIPTION EXAMPLE OF MOVIE "Ultra steel …"**

## DOC

Identifier = http://155.69.66.134:8000/Nhkvideo.mpg
Title = *Ultra-steel suspends the longest bridge in the world*
Publisher = NHK
Type = Doc
Description =        This movie is a scientific film about how Ultra-steel invented and what characteristics
                    it has to support the construction of the longest bridge in the world
Format =mpeg-1

## VIDEO

Identifier = http://155.69.66.134:8000/Nhkvideo.mpg
Title = *Ultra-steel suspends the longest bridge in the world*
Publisher = NHK
Type = Video
Description =        *This movie is a scientific film about how Ultra-steel invented and what characteristics*
                    *it has to support the construction of the longest bridge in the world*
Format =mpeg-1
Duration =  00:09:04:15
Rate = 30 frames/second

## MOVIE

Identifier = http://155.69.66.134:8000/Nhkvideo.mpg
Title =  *Ultra-steel suspends the longest bridge in the world*
Publisher = NHK
Type = Movie
Description =        *This movie is a scientific film about how Ultra-steel invented and what*
                    *characteristics it has to support the construction of the longest bridge in the world*
Format = MPEG-1
Duration = *00:09:04:15*
Rate = 30 frames/second
Movietype = scientific
Cast = Reporter, Tina Kawamura
Cast = Research Engineer, Dr. Toshihiko Takahashi

## SCENE #1

Identifier = http://155.69.66.134:8000/Nhkvideo.mpg#scene1
Title = *Ultra-steel suspends the longest bridge in the world*
Publisher = NHK
Type = Scene

Description =     *Air-shot of the Akashi-channel bridge*
Format = MPEG-1
Duration = *00:00:08:13*
Starttime= 00:00:00:00
endtime=00:00:08:13
Rate = 30 frames/second
object = bridge
Keyframe = http://155.69.66.134:8000/Nhkvideo.mpg#frame1

## FRAME 1.1

Identifier = http://155.69.66.134:8000/Nhkvideo.mpg#frame1
Title = *Ultra-steel suspends the longest bridge in the* world
Publisher = NHK
Type = Frame
Description =     Air-shot of the Akashi-channel bridge
Format = MPEG-1
object = bridge
color = Mixed

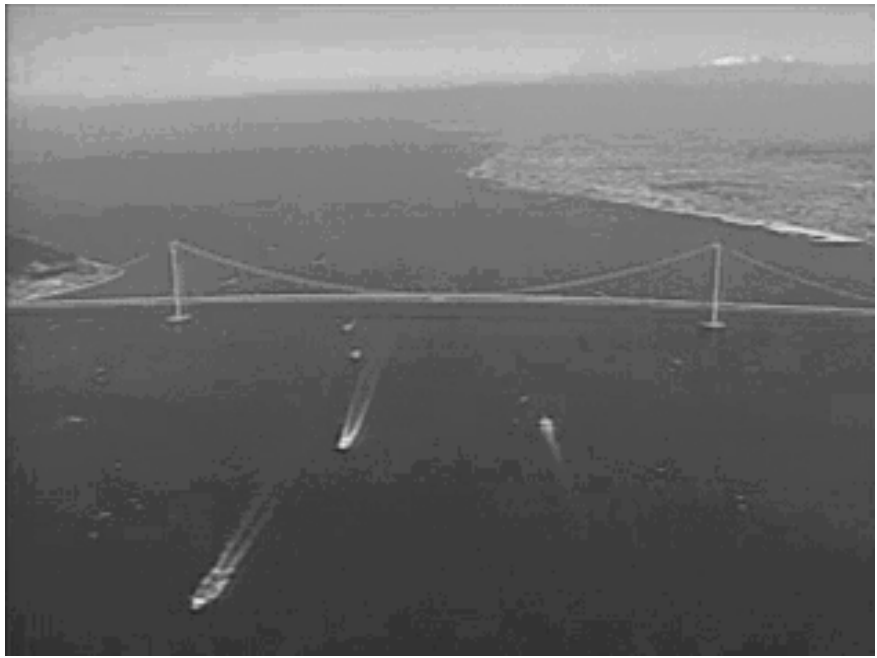

**Frame 1**

## Object #1.1.1

Identifier = http://155.69.66.134:8000/Nhkvideo.mpg#Object1.1.1
Title = *Akashi-channel bridge*
Publisher = NHK

Type = "Object"
Description =     Air-shot of the Akashi-channel bridge
color = N/A
position = N/A
motion = N/A
temporal = N/A

## SCENE  #2

Identifier = http://155.69.66.134:8000/Nhkvideo.mpg#scene2
Title = *Ultra-steel suspends the longest bridge in the world*
Publisher = NHK
Type = Scene
Description =     Panning-shot of the bridge
Format = MPEG-1
Duration = 00:00:08:06
Starttime=00:00:08:14
Endtime=00:00:16:20
Rate = 30 frames/second
object = *Akashi-channel bridge*
Keyframe = http://155.69.66.134:8000/Nhkvideo.mpg#frame.254

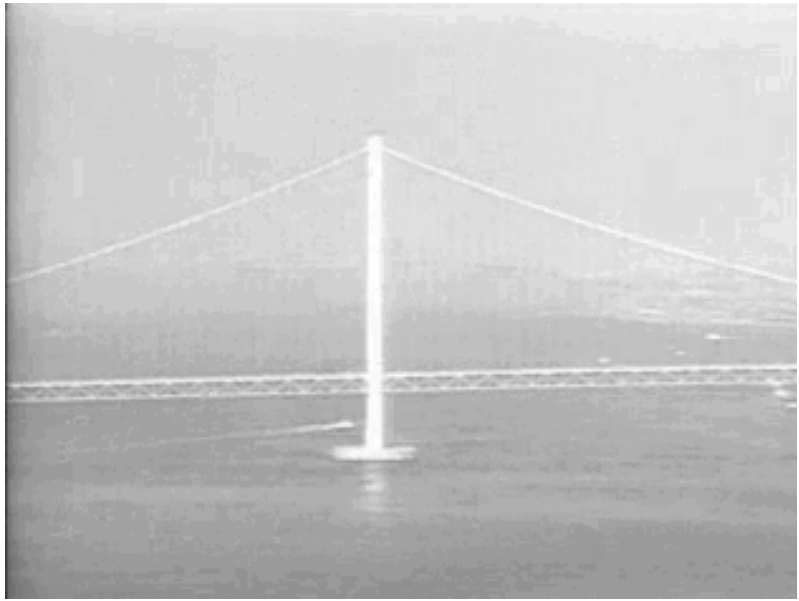

**Frame 254**

## FRAME #2.1

Identifier = http://155.69.66.134:8000/Nhkvideo.mpg#frame254
Title = *Ultra-steel suspends the longest bridge in the world*
Publisher = NHK
Type =Frame
Description =     Ralph Fiennes is putting his bag on the shoulder

Format = MPEG-1
object = Akashi-channel bridge
color = N/A

## OBJECT #2.1.1

Identifier = http://155.69.66.134:8000/Nhkvideo.mpg#Object2.1.1
Title = Akashi-channel bridge
Publisher = NHK
Type = Object
Description =     The length of the bridge is 4,000 meters. Its the longest suspension bridge in the
                world
color = N/A
position = N/A
motion = N/A
temporal = N/A

## SCENE 3

Identifier = http://155.69.66.134:8000/Nhkvideo.mpg#scene3
Title = *Ultra-steel suspends the longest bridge in the world*
Publisher = NHK
Type = Scene
Description =     Dolly-shot of cables of the Akashi-channel bridge
Format = MPEG-1
Duration = 00:00:14:00
Starttime=00:00:16:21
Endtime=00:00:30:21
Rate = 30 frames/second
object = Akashi-channel bridge, cable
Keyframe = http://155.69.66.134:8000/Nhkvideo.mpg#frame501

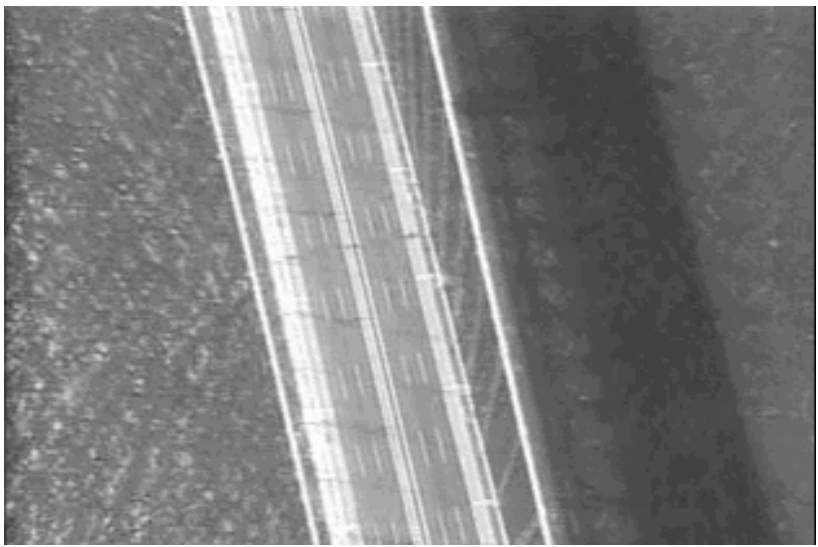

**Frame    501**

## FRAME #3.1

Identifier = http://155.69.66.134:8000/Nhkvideo.mpeg#frame501
Title = *Ultra-steel suspends the longest bridge in the world*
Publisher = NHK
Type = Frame
Description =                    Dolly-shot of cables of the Akashi-channel bridge
Format = MPEG-1
object = Akashi-channel bridge
color = N/A
texture = N/A

## OBJECT #3.1.1

Identifier = http://155.69.66.134:8000/Nhkvideo.mpg#Object3.1.1
Title = Akashi-channel bridge
Publisher = NHK
Type =Object
Description =      These two cables are suspending 150,000 ton of the weight. The cables are made of
                   special steel called Ultra-steel

color = N/A
position = N/A
motion = N/A
temporal = N/A